\documentclass[10pt]{iopart}
\usepackage{iopams,setstack}

\def\unity{\mbox{\small 1} \!\! \mbox{1}}

\begin{document}

\title[Quantum lithography \& parameter estimation]{Quantum
  lithography, entanglement and Heisenberg-limited parameter estimation}

\author{Pieter Kok,$^*$ Samuel L.\ Braunstein,$^{\dag}$ \\ 
  and Jonathan P.\ Dowling$^{\ddag}$}
\address{$^*$ Hewlett Packard Labs, Filton Road, Stoke Gifford, Bristol BS34 8QZ, UK}
\address{$^{\dag}$ Computer Science, York University, York YO10 5DD, UK}
\address{$^{\ddag}$ Quantum computing technologies group, Jet Propulsion Laboratory, California Institute of Technology, MS 126-347, 4800 Oak Grove Drive, Pasadena, CA}

\ead{pieter.kok@hp.com}

\begin{abstract}
 We explore the intimate relationship between quantum lithography,
 Heisenberg-limited parameter estimation and the rate of dynamical
 evolution of quantum states. We show how both the enhanced accuracy
 in measurements and the increased resolution in quantum lithography
 follow from the use of entanglement. Mathematically, the hyperresolution of
 quantum lithography appears naturally in the derivation of
 Heisenberg-limited parameter estimation. We also review recent
 experiments offering a proof of principle of quantum lithography, and
 we address the  question of state preparation and the fabrication of
 suitable  photoresists.
\end{abstract}

\submitto{\JOB}

%\maketitle

\bigskip

%\section{Introduction}

\noindent
An important branch of quantum mechanics is parameter
estimation. Heisenberg's uncertainty principle seems to prevent us
from determining a physical parameter such as a phase with infinite
precision, and it is therefore important to understand what are the limits
of the estimation process. It is also a branch of physics that
particularly interested H.A.\ Haus (e.g., Haus, 1995).

In this paper we consider the link between phase estimation and the
rate of dynamical evolution of quantum states. This will lead us to
the concept of quantum lithography and the question of the role of
entanglement. Finally, we will review some experiments that show the
viability of quantum lithography (at least in principle), and briefly
consider the generation of the necessary optical quantum states.

\section{Parameter estimation}

Consider a physical process that induces a relative phase $\varphi$ in
the state of a system. For example, one can send light through a
medium with an unknown index of refraction $n_r$ to induce a phase
shift $\varphi=2\pi L(n_r-1)/\lambda$, with $\lambda$ the wavelength
of the light and $L$ the length of the medium. This phase shift can be
measured relative to a reference beam in a Mach-Zehnder
interferometer.  When we assume that $L$ and $\lambda$ are known with
very high precision, we can infer the index of refraction with an
accuracy that is proportional to the error in the phase. Many
precision measurements can be reformulated in terms of a phase
measurement, with the search for gravitational waves as one of the
most urgent cases (Fritschel, 1998). It is therefore important to have
definite bounds on the error in the phase, given a particular set of
resources. We should realize, however, that there are two distinct
questions we can ask about the uncertainty in the phase.

First, there is the relative phase difference $\delta\varphi$ that 
renders two quantum states $|\psi(0)\rangle$ and $|\psi(\delta\varphi)\rangle$ 
distinguishable, i.e., 
\begin{equation}\label{dist}
  \langle\psi(0)|\psi(\delta\varphi)\rangle = 0\; . 
\end{equation}
Here, we assume that the dependency on $\delta\varphi$ actually does render
$|\psi(\delta\varphi)\rangle$ orthogonal to $|\psi(0)\rangle$. We can
then minimize $\delta\varphi$ to satisfy Eq.\ (\ref{dist}). This
depends on the characteristics of $|\psi(0)\rangle$ and
$|\psi(\delta\varphi)\rangle$. Thus, $\delta\varphi$ is a measure of
the  dynamical rate of the (unitary) evolution.

A second question we can ask, is how well one can measure
$\varphi$ given a state $|\psi(\varphi)\rangle$. That is, we want an
expression for the error $\Delta\varphi$ in the phase. Since there is no
phase operator, we usually measure a suitably chosen (Hermitian)
operator $\hat{X}$ such that 
\begin{equation}\label{est}
  \Delta\varphi = \frac{\Delta\hat{X}}{|d\langle\hat{X}\rangle/d\varphi|}\; .
\end{equation}
Mathematically, these two problems are very similar, since measuring
$|\psi(\varphi)\rangle$ typically involves interference with some
other state $|\psi(\varphi')\rangle$. Choosing  situations where
$\delta\varphi = \Delta\varphi$ allows us to derive  bounds on the
fundamental quantum limits of parameter estimation  ($\Delta\varphi$)
by calculating the maximum rate of dynamical evolution
($\delta\varphi$).

It was recognized early on by Mandelstam and Tamm in 1945 that the 
uncertainty relation between time and energy for a given frequency 
$\omega$ can be used to prove the following inequality (Mandelstam and 
Tamm, 1945):
\begin{equation}
 \Delta\varphi = \omega\Delta t \geq \frac{\pi}{2}
 \frac{\hbar\omega}{\Delta E}\; ,
\end{equation}
where $t$ is the time it takes to evolve from the initial state to the
next orthogonal state, $\hbar\omega$ is the energy difference between
the two orthogonal states and we have used $\delta\varphi =
\Delta\varphi$. Furthermore, $\Delta E$ is the uncertainty in energy.
This inequality can be interpreted as follows: the minimum phase
difference $\Delta\varphi$ that can be detected is inversely
proportional to the normalized energy spread of the state 
$|\psi\rangle$ (with $(\Delta E)^2 = \langle\psi|E^2|\psi\rangle -
\langle\psi|E|\psi\rangle^2$). The normalization is given by
$\hbar\omega$. The factor $\pi/2$ has a geometric interpretation that
is highly relevant for this paper: The minimum distance between
two points that can be distinguished by a wave is given by the
distance between a crest and the adjacent trough of the wave. This 
(normalized) distance of $\pi/2$ is also known as the {\it Rayleigh limit}.

However, it became clear that the Mandelstam-Tamm (MT) inequality is not
applicable to situations involving certain pathological states (Shapiro 
{\em et al}., 1989; Braunstein {\em et al}., 1992; Uffink, 1993). When 
coherent (classical) states with $\Delta E = \hbar\omega 
\sqrt{\langle n\rangle}$ are used, the MT-inequality yields 
the Poissonian error
\begin{equation}
 \Delta\varphi \geq \frac{\pi}{2} \frac{1}{\sqrt{\langle n\rangle}}\; ,
\end{equation}
where $\langle n\rangle$ is the average number of quanta in the coherent 
state. The indication was that phase estimation could not improve 
beyond the so-called Heisenberg limit (Ou, 1996):
\begin{equation}\label{ou}
 \Delta\varphi \gtrsim \frac{1}{\langle n\rangle}\; .
\end{equation}
But there exist states with finite average energy $E$ and unbounded
$\Delta E$. According to the MT-inequality, such states could have a
precision that is proportional to, for example, $\langle n\rangle^{-2}$. 

If there are states for which the MT-inequality does not give a
saturated bound, then the next question is whether an additional bound
can be derived. In 1998, Margolus and Levitin did just that (Margolus 
and Levitin, 1998). They showed that
\begin{equation}
 \Delta t \geq \frac{\pi}{2} \frac{\hbar}{E}\; ,
\end{equation}
which for narrow-band schemes around frequency $\omega$ with $E =
\hbar\omega \langle n \rangle$ yields
\begin{equation}
 \Delta\varphi \geq \frac{\pi}{2} \frac{1}{\langle n \rangle}\; .
\end{equation}
This inequality therefore restricts the precision in a phase
measurement by the {\em average energy}, rather than the energy
spread.  The pathological states mentioned above turn out to have
sufficiently  small average energies so that Eq.~(\ref{ou}) is still
satisfied. Recent examples of applied phase estimation include frequency
measurements (Bollinger, 1996), the quantum gyroscope (Dowling, 1998),
quantum positioning and clock synchronization (Giovannetti et al.,
2001), and length and weak force  measurements using coherent optical 
states (Ralph, 2002; Munro {\em et al.}, 2002).

\section{The role of entanglement in parameter estimation}

Suppose we want to estimate a parameter $\varphi$ by measuring a 
particular observable. Construct the state 
\begin{equation}\label{instate}
  |\varphi\rangle = \frac{1}{\sqrt{2}}\left( |0\rangle + e^{i\varphi}
   |1\rangle \right)\; ,
\end{equation}
where the basis $\{ |0\rangle,|1\rangle\}$ is chosen in some
convenient way determined by the physics of estimating $\varphi$ with
the choice of observable as $\sigma_x = |0\rangle\langle 1| + |1\rangle\langle
0|$. Then the expectation value of $\sigma_x$ is given by
$\langle\varphi|\sigma_x|\varphi\rangle = \cos\varphi$. When we repeat
this experiment $N$ times, we obtain
\begin{equation}
  \langle\sigma_x^{N}\rangle=_1\langle\varphi|\ldots\, _N\!\langle\varphi|
  \,\overset{N}{\underset{k=1}{\mbox{\Large $\oplus$}}}\,
  \sigma_x^{(k)}|\varphi\rangle_1 \ldots |\varphi\rangle_N = N
  \cos\varphi\; .
\end{equation}
Furthermore, we know that $\sigma_x^2=\unity$, and the variance of
$\sigma_x$ given $N$ samples is readily computed to be
$(\Delta\sigma_x)^2 = N(1-\cos^2\varphi)=N\sin^2\varphi$.
According to estimation theory
\begin{equation}
  \Delta\varphi =
  \frac{\Delta\sigma_x}{|d\langle\sigma_x^{N}\rangle/d\varphi|}\; .
\end{equation}
The standard variance in the parameter $\varphi$ after $N$ trials
is thus given by 
\begin{equation}
  \Delta\varphi_{\rm st} = \frac{\sqrt{N}\,\sin\varphi}{N
  \sin\varphi} = \frac{1}{\sqrt{N}}\; .
\end{equation}
In other words, the uncertainty in the phase is inversely proportional
to the square root of the number of trials. This is the classical 
Poissonian error in the phase.

With the help of quantum entanglement we can achieve the Heisenberg 
limit of $1/N$. Consider an entangled input state on $N$ systems:
\begin{equation}\label{entstate}
  |\varphi_N\rangle = \frac{1}{\sqrt{2}}\left(
   |0,\ldots,0\rangle_{1\ldots N} + e^{iN\varphi}
   |1,\ldots,1\rangle_{1\ldots N} \right)\; .
\end{equation}
The relative phase $e^{iN\varphi}$ can be obtained by a unitary
evolution $|1\rangle \rightarrow e^{i\varphi} |1\rangle$ and $|0\rangle 
\rightarrow |0\rangle$, thus yielding the required factor. When we 
suggestively write $|{\mathbf 0}\rangle = |0,\ldots,0\rangle_{1\ldots N}$ and 
$|{\mathbf 1}\rangle = |1,\ldots,1\rangle_{1\ldots N}$, then the state becomes
\begin{equation}\label{inst}
  |\varphi_N\rangle = \frac{1}{\sqrt{2}}\left( |{\mathbf 0}\rangle +
  e^{iN\varphi} |{\mathbf 1}\rangle \right)\; .
\end{equation}
This is mathematically equivalent to a single (nonlocal!) system with 
a relative phase shift of $N\varphi$. In order to measure this phase, 
we need to measure the (nonlocal) observable $\Sigma_N$:
\begin{equation}
 \Sigma_N = |{\mathbf 0}\rangle\langle{\mathbf 1}| +
 |{\mathbf 1}\rangle\langle{\mathbf 0}|\; .
\end{equation}
This yields
\begin{equation}\label{entav}
  \langle\varphi_N|\Sigma_N|\varphi_N\rangle = \cos N\varphi\; .
\end{equation}
As before, we obtain
\begin{equation}
    \Delta\varphi_q = 
    \frac{\Delta\Sigma_N}{|d\langle\Sigma_N\rangle/d\varphi|} =
    \frac{1}{N} \frac{\sin N\varphi}{\sin N\varphi} = \frac{1}{N}\; . 
\end{equation}
Here we see that the precision in $\varphi$ is increased by a factor
$\sqrt{N}$ over the standard noise limit when we exploit quantum
entanglement. 

When the loss of subsystems is considered (for example, when
$n$ out of $N$ photons fail to arrive), the entangled state in Eq.\
(\ref{entstate}) becomes separable and mixed:
\begin{equation}
  \rho_{N-n} = \frac{1}{2}|0\rangle_{1\ldots N-n}\langle 0| +
  \frac{1}{2}|1\rangle_{1\ldots N-n}\langle 1|\; .
\end{equation}
Not only is it separable, there is also no information about $\varphi$
in this state. This state can therefore not be used for the parameter
estimation at all.

One way to circumvent this practical difficulty is to use separable
states and non-local observables. When $N-n$ out of $N$ systems
arrive, the experimenter chooses to measure $\Sigma_{N-n}$ instead of
$\Sigma_N$. This way, the parameter $\varphi$ is estimated with an
enhanced precision proportional to $1/(N-n)$. The price to pay is that
often the measurement of $\Sigma_{N-n}$ will fail because it does not
span the complete state space. It must also be noted that this trick
depends on the physical implementation, and  cannot always be applied.

\section{Quantum lithography}

One particular physical implementation of this process involves light,
and this will lead us to the main topic of this paper. Suppose that
the state in Eq.~(\ref{inst}) is physically implemented by a so-called
two-mode {\em Noon state} (Kok {\em et al}., 2002):
\begin{equation}
 |\varphi_N\rangle = \frac{1}{\sqrt{2}} \left( |N,0\rangle +
 |e^{iN\varphi} 0,N\rangle \right)\; ,
\end{equation}
where $|k\rangle$ is an $k$-photon state. When we choose $\Sigma_N$ to
be
\begin{equation}
 \Sigma_N = |N,0\rangle\langle 0,N| + |0,N\rangle\langle N,0|\; ,
\end{equation}
then the uncertainty in the phase is again given by $\Delta\varphi =
1/N$.  However, there is also something else going on \ldots

When we calculate the expectation value of $\Sigma_N$, we find that it
varies as $\cos N\varphi$. That is, the geometric distance between the
two points that can be distinguished according to the Rayleigh  limit
is now {\em $N$ times smaller than in the classical case!} As a
consequence, we can in principle read and write much smaller features
with this technique. This is  called quantum lithography (Boto {\em et
al}., 2000), and we will now  give a full description of this
phenomenon in terms of the quantum interference of two optical modes.

\subsection{Interference on a surface and the Rayleigh limit}

Suppose two plane waves characterised by $\vec{k}_1$ and $\vec{k}_2$
hit a  surface at an angle $\theta$ from the normal vector. The wave
vectors are  given by
\begin{equation}\label{planewave}
 \vec{k}_1 = k(\cos\theta,\sin\theta) \quad\mbox{and}\quad \vec{k}_2 =
 k(\cos\theta,-\sin\theta)\; ,
\end{equation}
where we assume $|\vec{k}_1|=|\vec{k}_2|=k$. The wave number $k$ is
related to  the wavelength of the light according to $k=2\pi/\lambda$.

In order to find the interference pattern in the intensity $I$, we sum
the  two plane waves at position $\vec{r}$ at the amplitude level:
\begin{equation}
 I(\vec{r}) \propto
 \left|e^{i\vec{k}_1\cdot\vec{r}}+e^{i\vec{k}_2\cdot\vec{r}} \right|^2
 = 4\cos^2\left[ \frac{1}{2}(\vec{k}_1 - \vec{k}_2)\cdot\vec{r}
 \right]\; .
\end{equation}
When we calculate the inner product $(\vec{k}_1 -
\vec{k}_2)\cdot\vec{r}/2$  from Eq.\ (\ref{planewave}) we obtain the
expression
\begin{equation}\label{cos2}
 I(x) \propto \cos^2(kx\sin\theta)
\end{equation}
for the intensity along the substrate in direction $x$.

As we saw above, the Rayleigh limit is given by the minimal resolvable
feature size $\Delta x$ that corresponds to the distance between an
intensity maximum and an adjacent minimum. From Eq.\ (\ref{cos2}) we
obtain
\begin{equation}
 k\Delta x\sin\theta = \frac{\pi}{2}\; .
\end{equation} 
This means that the maximum resolution is given by
\begin{equation}
 \Delta x = \frac{\pi}{2k\sin\theta} =
 \frac{\pi}{2\left(\frac{2\pi}{\lambda} \sin\theta\right)} =
 \frac{\lambda}{4\sin\theta}\; .
\end{equation}
The maximum resolution is therefore proportional to the wavelength and
inversely proportional to the  sine of the angle between the incoming
plane waves and the normal. The  resolution is thus maximal ($\Delta
x$ is minimal) when $\sin\theta=1$, or  $\theta=\pi/2$. This is the
limit of grazing incidence. The classical  diffraction limit is
therefore $\Delta x = \lambda/4$. Note that this  derivation does not
use the approximation $\sin\theta\simeq\theta$, which  is common when
considering diffraction phenomena.

\subsection{Surpassing Rayleigh's diffraction limit}

So how does quantum lithography work? In the limit of grazing
incidence, we let the two counter-propagating  light beams $a$ and $b$
be in the combined entangled state of $N$ photons
\begin{equation}\label{n00n}
 |\psi_N\rangle_{ab} = \left( |N,0\rangle_{ab} + e^{iN\varphi} |0,N\rangle_{ab}
 \right) / \sqrt{2}\; ,
\end{equation}
where $\varphi=kx$, with $k=2\pi/\lambda$.
We define the mode operator $\hat{e} = (\hat{a}+\hat{b})/\sqrt{2}$ and its 
adjoint $\hat{e}^{\dagger} = (\hat{a}^{\dagger}+\hat{b}^{\dagger})/\sqrt{2}$. 
The deposition rate $\Delta$ on the substrate is then given by (Boto
{\em et al}., 2002):
\begin{equation}\label{delta}
 \Delta_N = \langle\psi_N|\hat{\delta}_N|\psi_N\rangle\qquad\mbox{with}\qquad
 \hat{\delta}_N=\frac{(\hat{e}^{\dagger})^N \hat{e}^N}{N!}\; ,
\end{equation}
i.e., we look at the higher-order moments of the electric field operator on 
the substrate.
The deposition rate $\Delta$ scales with the $N^{\rm th}$ power of
intensity.  Leaving the  substrate exposed for a time $t$ to the light
source will result in an  exposure pattern $P(\varphi)=\Delta_N
t$. After a straightforward calculation  we see that
\begin{equation}\label{deprate}
 \Delta_N \propto (1 + \cos N\varphi)\; .
\end{equation}
We interpret this as follows. A path-differential phase-shift
$\varphi$ in  light beam $b$ results in a displacement $x$ of the
interference pattern on  the substrate. Using two classical waves, a
phase-shift of $2\pi$ will return  the pattern to its original
position. However, according to  Eq.~(\ref{deprate}), one cycle is
completed after a shift of $2\pi/N$. This  means that a shift of
$2\pi$ will displace the pattern $N$ times. In other  words, we have
$N$ times as many maxima in the interference pattern. These will
be closely spaced, yielding an effective Rayleigh resolution of
$\Delta x = \lambda/4N$, a factor of $N$ below the classical
interferometric result of  $\Delta x=\lambda/4$.

We have also shown that quantum lithography can be used to create
arbitrary patterns in one dimension, rather than just closely spaced
lines (Kok {\em et al}., 2001). This can be achieved by using
superpositions of the more complicated state
\begin{equation}\label{nm}
 |\psi_{Nm}\rangle = \frac{1}{\sqrt{2}} \left( e^{im\varphi}
 |N-m,m\rangle + e^{i(N-m)\varphi} |m,N-m\rangle 
 \right)\; .
\end{equation}
There are two fundamentally different ways states of the form in
Eq.~(\ref{nm}) can be superposed. We can either sum over the photon 
number $N$:
\begin{equation}\label{n}
 |\Psi_m\rangle = \sum_{n=0}^N \alpha_n |\psi_{nm}\rangle\; ,
\end{equation}
with $\alpha_n$ complex coefficients, or we can sum over the photon 
distribution $m$:
\begin{equation}\label{bjork}
 |\Psi_N\rangle = \sum_{m=0}^{\lfloor N/2\rfloor} \alpha_m|\psi_{Nm}\rangle\; ,
\end{equation}
where $\lfloor N/2\rfloor$ denotes the largest integer $l$ with $l\leq N/2$ and
$\alpha_m$ again the complex coefficients. Every branch in this superposition
is an $N$-photon state.

These techniques not only allow us to create arbitrary one-dimensional
patterns, but extensions to four-mode states also facilitate
two-dimensional patterns (Kok {\em et al}., 2001). Furthermore, by
choosing the $\alpha_m$ coefficients of Eq.~(\ref{bjork}) in a special
way, Bj\"ork {\em et al}.\ showed that one can construct a
subwavelength resolution pixel state. This state can subsequently be
used to illuminate a surface in order to etch an arbitrary pattern 
(Bj\"ork {\em et al}., 2001).

\subsection{Demonstration of quantum lithography and the preparation of states}

The principle of quantum lithography has been demonstrated using
two-photon path-entangled states generated by parametric
down-conversion (d'Angelo et al., 2001). In this experiment, the
photon pairs are created in a BBO crystal ($\beta$-${\rm BaB}_2{\rm
O}_4$). Immediately behind the crystal, a double-slit aperture is
placed, which blocks most pairs. However, when two photons do get
through, they are extremely likely to have passed through the same 
slit. 

This way, the state of the light field just after the slits is
approximately $|2,0\rangle_{AB} + |0,2\rangle_{AB}$, where $A$ and $B$
denote the two slits. This state is of the form of Eq.~(\ref{n00n}),
and can therefore be used to beat the Rayleigh limit. By scanning the
output field with a set of two detectors, the diffraction pattern
conditioned on a two-fold detector coincidence was mapped out. This
pattern was twice as narrow as the single-photon diffraction
pattern, and the principle of sub-Rayleigh resolution due to
two-photon quantum lithography was thereby demonstrated. Subsequent
experiments with photon pairs have confirmed these findings (Edamatsu
{\em et al}., 2002; Shimizu {\em et al}., 2002). 

However, in order for quantum lithography to work, we also need a
suitable photo-resist. In other words, we need a material that is
sensitive to multi-photon events in order to etch images on its
surface. Last year, Korobkin and Yablonovich exposed commercial
photographic film to photon pairs, and produced the coveted two-fold
resolution enhancement (Korobkin and Yablonovitch, 2002). Since this
experiment includes both the state preparation and the imaging
process, it qualifies as the first {\em complete} demonstration of
quantum lithography. 

Shortly after the experiment of d'Angelo {\em et al}., Cataliotti {\em
et al}., performed frequency measurements using multiphoton Raman
transitions in Rubidium atoms that are confined in an optical dipole
trap (Cataliotti {\em et al}., 2001). The multiphoton events of up to
fifty photons resulted in a spectral width that is below the Fourier
limit. Although not exactly quantum lithography, this experiment
strongly suggests that the {\em quantum Rayleigh limit} of
$\lambda/4N$ is correct.

\bigskip

Another essential ingredient for quantum lithography is the generation
of the required quantum states of the light field. The states in
Eqs.~(\ref{nm}) and (\ref{n}) are very complicated, and it is not
quite clear how they can be generated efficiently without large Kerr
nonlinearities (Gerry and Campos, 2001). The production of $N$-photon 
entangled states conditioned on {\em non}-detection was proposed
(Fiur\'a{\v s}ek, 2002), as well as the creation of Noon-states based
on single-photon detection (Lee {\em et al}., 2002; Kok {\em et al}.,
2002; Gerry {\em et al}., 2002).

\section{Conclusions}

We have shown that quantum lithography and Heisenberg-limited
parameter estimation are two manifestations of the same principle:
Instead of many separate measurements of $\varphi$ that lead to the
shot-noise limit (in, for example, an experiment using $N$ trials with the
single-photon path-entangled state $|1,0\rangle + e^{i\varphi}|0,1\rangle$),
entangling the resources of these $N$ measurements to conduct a single-shot
experiment (e.g., using the $N$-photon Noon-state $|N,0\rangle +
|0,N\rangle$) can reduce the noise to the Heisenberg limit. Similarly,
instead of Rayleigh limited single-photon diffraction patterns, we can
use $N$-photon entangled states to increase the resolution by a factor
$N$. Mathematically, the hyperresolution of quantum lithography
appears naturally in the derivation of Heisenberg-limited parameter
estimation.

\section*{Acknowledgements}

Part of the research described in this paper was carried out at the
Jet Propulsion Laboratory, California Institute of Technology, under a
contract with the National Aeronautics and Space Administration
(NASA). P.K.\ was supported by the European Union RAMBOQ project, and
S.L.B.\ currently holds a Royal Society-Wolfson research merit award.

\section*{References}

\begin{harvard}
 \item[] d'Angelo M, Chekhova M V and Shih Y 2001 {\it Phys.\ Rev.\ Lett.} 
   {\bf 87} 013602
 \item[] Bj\"ork G, Sanchez-Soto L L and Soderholm J 2001 
   {\it Phys.\ Rev.\ Lett.} {\bf 86} 4516
 \item[] Bollinger J J, Itano W M, Wineland D J and Heinzen D J 1996 
   {\it Phys.\ Rev.\ A} {\bf 54} R4649
 \item[] Boto A N, Kok P, Abrams D S, Braunstein S L, Williams C P and 
   Dowling J P 2000 {\it Phys.\ Rev.\ Lett.} {\bf 85} 2733
 \item[] Braunstein S L, Lane A S and Caves C M 1992 {\it Phys.\ Rev.\ 
   Lett.} {\bf 69} 2153
 \item[] Cataliotti F S, Scheunemann R, H\"ansch T W and Weitz M 2001
   {\it Phys.\ Rev.\ Lett.} {\bf 87} 113601
 \item[] Dowling J P 1998 {\it Phys.\ Rev.\ A} {\bf 57} 4736
 \item[] Edamatsu K, Shimizu R and Itoh T 2002
   {\it Phys.\ Rev.\ Lett.} {\bf 89} 213601
 \item[] Fiur\'a{\v s}ek J 2002 {\it Phys.\ Rev.\ A} {\bf 65} 053818
 \item[] Fritschel P, Gonzalez G, Lantz B, Saha P and Zucker M 1998 
   {\it Phys.\ Rev.\ Lett.} {\bf 80} 3181
 \item[] Gerry C C, Benmoussa A and Campos R A 2002 {\it Phys.\ Rev.\ A} 
   {\bf 66} 013804
 \item[] Gerry C C and Campos R A 2001 {\it Phys.\ Rev.\ A} {\bf 64} 063814
 \item[] Giovannetti V, Lloyd S and Maccone L 2001 {\it Nature} {\bf 412} 417
 \item[] Haus H A 1995 {\it J.\ Opt. Soc.\ Am.\ B} {\bf 12} 2019
 \item[] Kok P, Boto A N, Abrams D S, Williams C P, Braunstein S L and 
   Dowling J P 2001 {\it Phys.\ Rev.\ A} {\bf 63} 063407
 \item[] Kok P, Lee H and Dowling J P 2002 {\it Phys.\ Rev.\ A} {\bf 65} 052104
 \item[] Korobkin D V and Yablonovitch E 2002 {\it Opt.\ Eng.} {\bf 41} 1729
 \item[] Lee H, Kok P, Cerf N J and Dowling J P 2002 {\it Phys.\ Rev.\ A} 
   {\bf 65} 030101
 \item[] Mandelstam L and Tamm I 1945 {\it J.\ Phys.} (USSR) {\bf 9} 249
 \item[] Margolus N and Levitin L B 1998 {\it Phys.\ D} {\bf 120} 188
 \item[] Munro W J, Nemoto K, Milburn G J and Braunstein S L 2002 
   {\it Phys.\ Rev.\ A} {\bf 66} 023819
 \item[] Ou Z Y 1996 {\it Phys.\ Rev.\ Lett.} {\bf 77} 2352
 \item[] Ralph T C 2002 {\it Phys.\ Rev.\ A} {\bf 65} 042313
 \item[] Shapiro J H, Shepard S R and Wong N C 1989 {\it Phys.\ Rev.\ 
   Lett.} {\bf 62} 2377
 \item[] Shimizu R, Edamatsu K and Itoh T 2002 {\it Pram.\ J.\ Phys.} 
   {\bf 59} 165
 \item[] Uffink J 1993 {\it Am.\ J.\ Phys.} {\bf 61} 935
\end{harvard}

\end{document}